\documentclass[pra,twocolumn,aps,showpacs,amsmath,amssymb]{revtex4-1}

\usepackage{color}
\usepackage{bm}
\usepackage{graphicx}
\usepackage{braket}

\usepackage[plainpages=false,pdfpagelabels,colorlinks=true,linkcolor=red,urlcolor=blue,citecolor=blue,pdftitle={Title},pdfauthor={},pdfdisplaydoctitle=true,pdfduplex=DuplexFlipLongEdge]{hyperref}

\def\Tr{\mbox{Tr}\,}

\newcommand\ba{\begin{eqnarray}}
\newcommand\ea{\end{eqnarray}}
\newcommand\be{\begin{equation}}
\newcommand\ee{\end{equation}}

\begin{document}

\title{Finite-size scalings in measurement-induced dynamical phase transition}
\author{Ranjan Modak$^{1,3}$, Debraj Rakshit$^{2,3}$ and Ujjwal Sen$^2$} 
\affiliation{$^1$ Department of Physics, Indian Institute of Technology Tirupati, Tirupati 517506, India}
\affiliation{$^2$ Harish-Chandra Research Institute, HBNI, Chhatnag Road, Jhunsi, Allahabad 211 019, India}
\affiliation{$^3$ Department of Physics,Institute of Science, BHU, Varanasi-221005, India}

\begin{abstract}
Repetitive measurements can cause  freezing of dynamics of a quantum state, which is known as quantum Zeno effect. We consider an interacting one-dimensional fermionic system and study the fate of the many-body quantum Zeno transition if the system is allowed to evolve repetitively under the unitary dynamics, followed by a measurement process. Measurement induced phase transitions can be accessed by tuning a suitably defined parameter representing measurement strength (frequency). We use different diagnostics, such as long-time evolved entanglement entropy, purity and their fluctuations in order to characterize the transition. We further perform a finite size scaling analysis in order to detect the transition points and evaluate associated scaling exponents via an unbiased numerical strategy of cost function minimization, which provides a platform to compare finite-size scaling ans{\"a}tze proposed previously in context of many-body Zeno transition.

\end{abstract}
\maketitle

\section{Introduction}
Quantum measurement is a crucial element of quantum theory, and in particular in the fledgling field of quantum technologies. It finds important roles, e.g., in resource theory \cite{Chou05,Roch14,Poop05,Sadhukhan17} and quantum computation \cite{Raussendorf,Sorensen03,Brigel09}, and leads to fascinating quantum phenomena, such as the quantum Zeno effect \cite{Koshino05,Wiseman09}. In the quantum Zeno effect, frequent wave function collapses, induced by repetitive measurements, tend to stagnate the time evolution of a quantum mechanical system. Since its initial conceptualization \cite{Misra77}, there has been a substantial number of studies on the quantum Zeno phenomenon in various systems \cite{KOfman00,Facchi02,Maniscalco08,Facchi10,Militello11,Raimond12,Wang13,Stannigel}. In particular, quantum Zeno effects in quantum many-body systems have drawn significant attention in recent times \cite{Stannigel,Froml19,Froml20,Syassen08,Biella19}.

The concept of entanglement entropy from quantum information theory often leads towards building a natural information-theoretic understanding of quantum many-body phenomenology \cite{Amico08,Eisert10}. Importantly, entanglement entropy is now accessible in cold-atom experiments \cite{Islam15,Kaufman16}. Recent works have pointed out a phase transition in the form of  (sub)-volume-law to area-law (in the quantum Zeno regime) scaling of entanglement entropy in presence of quantum measurements. In general, a weakly entangled quantum state under a entanglement-rich driving Hamiltonian accumulates additional (quantum) correlations between its sub-parts during the course of unitary evolution. Accordingly, the bipartite entanglement entropy of a long-time evolved state turns out to be of the order of the system size. However, such growth of the entanglement entropy is inhibited in presence of the quantum measurements. In particular, long-time evolved entanglement entropy becomes of the order of the boundary area once the projective measurements are \emph{frequent enough}. Hence, measurement frequency acts as a control parameter that induces volume- to area-law  transition, and consequently, leads the system in the quantum Zeno phase.
\begin{figure}
    \centering
    \includegraphics[width=0.45\textwidth]{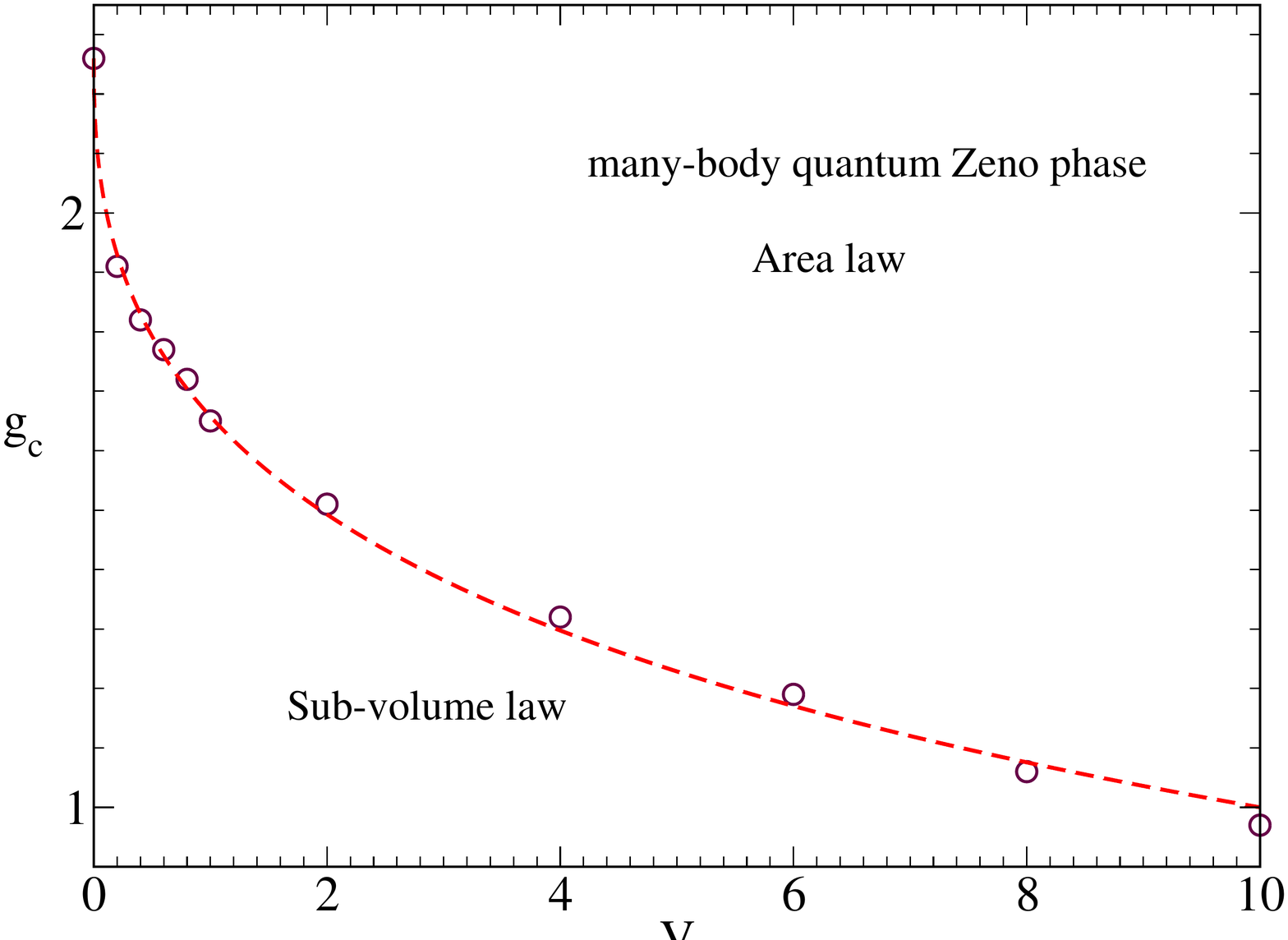}
    \caption{Phase diagram for quantum many-body Zeno transition as a function of 
    measurement strength $g$ and interaction strength $V$.  The dashed line corresponds to $g_c=2.26\exp(-0.31 V^{0.42})$, which is the best fit lines obtained the numerical data. }
    \label{fig0}
\end{figure}

Measurement-induced phase transitions have been reported in many-body context, such as in random circuits \cite{Chen18,Skinner19,Li19,Gullans19,Gullans20,Jian20,Bao20,Choi20,Fan20}, spin chains \cite{Dhar16,Turkeshi20a,Turkeshi20b,Maimbourg21}, and also very recently in long-range model \cite{Minato21}. Particularly, the nature of the transitions and their scalings hold special interest \cite{Zabalo20,Zabalo21}. A number of recent studies suggest a conformal critical point in the measurement induced transition \cite{Fuji20,Alberton21,Chen20,Buchhold21,Turkeshi21}, which differs from the treatment via algebraic scaling ansatz initially proposed by Li \emph{et. al.} \cite{Chen18,Li19}. The many-body complexity, along-with the probabilistic description of quantum mechanics, makes it difficult to cross-examine the quality of the proposed scaling ans{\" a}tze. This work adopts an unbiased means via so-called cost function minimization approach in order to closely inspect the scaling associated with measurement-induced phase transition in a one-dimensional interacting fermionic system. 

The generic feature of measurement-induced phase transition in entanglement entropy can be identified in the system under study, as one may expect by the virtue of universality. We, in particular, aim towards performing an accurate finite-size scaling analysis at finite interactions and report the phase diagram from there-off. In order to do so, the scaling analyses are performed within a general framework via several complementary measures, that includes, apart from entanglement entropy, another quantity - purity of the reduced density matrix. Purification plays important roles in several quantum-information theoretic ideas and applications \cite{Nielsen11,Wilde13,Kleinmann06,Bera16,Fang20}. Particularly, there has been a steady experimental development towards projective measurement based  purification schemes in context of quantum computers \cite{Hume07,Jiang09,Ofek16,Nakajima19}. Similar to the entanglement entropy, there is a measurement-induced dynamical phase transition in  linearized purity as well. A pure product state under weak measurement strength is expected to gain a considerable amount of entanglement during unitary evolution, ensuring a close to maximally mixed reduced density matrix. In this regime mixedness rapidly grows with system size. Enhanced measurement strength implies an enhanced purification of the reduced density matrix of the system. Beyond the transition point the purity obeys the area law in the quantum Zeno phase.

The results are bench-marked following two key steps. First, the critical points are first coarse-grained from the extremums of the entanglement fluctuations. We then resort to a newly introduced unbiased numerical recipe for extracting the fine-grained  transition points \cite{lev1}. This is done via cost function approach in order to estimate the quality of the finite-size collapse of the data corresponding to the transition indicators under study, i.e. entanglement entropy and purity. It is worth mentioning that all the indicators offer a consistent picture over the extracted information regarding the transition points and the scaling exponents over a wide range of interactions under consideration. Higher quality in the data collapse confirmed by the cost function minimization approach clearly asserts the conformal nature of the critical point. Moreover, we investigate the robustness of our scaling analysis by considering different initial states. 

The rest of the paper is arranged as follows. Sec.~II introduces the model and measurement protocols are presented.  Section II also presents the phase-diagram showing the measurement-induced Zeno transition line and the details  of involved methodologies are provided in the following sections.  The entanglement dynamics and related scaling analysis are described in Sec.~III. Section IV put forward the analysis related to purity. The conclusions are drawn in Sec.~V. In Sec. VII, Appendix I contains a short discussion over the choice of initial state.

\section{Model and Protocols}
We consider a one-dimensional fermionic lattice of length $L$ described by following Hamiltonian:
\begin{eqnarray}
\hat{H}&=&-\sum_{j}(\hat{c}^{\dag}_j\hat{c}_{j+1}+\text{H.c.}) +V\sum_j\hat{n}_j\hat{n}_{j+1}, \nonumber 
\label{nonint_model}
\end{eqnarray}
where $\hat{c}^{\dag}_j$  ( $\hat{c}_j$) is the fermionic creation (annihilation) operator at site $i$, $\hat{n}_j =\hat{c}^{\dag}_j\hat{c}_{j}$ is the number operator, and $V$ is the interaction strength.  We use periodic boundary condition to reduce the finite-size effects. %The Hamiltonian ~\eqref{nonint_model} can be mapped on the spin $1/2$ XXZ chain, which is integrable and exactly solvable by the Bethe ansatz ~\cite{Takahashi}.
We consider following stochastic protocol: A many-body quantum state evolves under the driving Hamiltonian $\hat{H}$ for a time interval $\Delta t$, and it is then followed by measurement carried out simultaneously on each $L$ lattice sites. For a given choice of an initial state, a long-time evolved steady state is achieved by repeating the protocol for several times. Let, $\psi(t)$ is the evolved state at the time $t$. The evolved state at time $t+\Delta t$ under one-time action can be written as $|\psi(t+\Delta t)\rangle=\hat{M}\hat{U}|\psi(t)$, 
where
$\hat{U}=e^{-i\hat{H}\Delta t}$ 
and henceforth, $t$ will be in units of $\hbar$. The operator $\hat{M}$ associated with projective measurements is given by $\hat{M}=\otimes_{j=1}^{L}\hat{M}^{r}_j$ with $r=0$,1, where
\begin{eqnarray}
\hat{M}^{0}_j&=&\hat{n}_j+\sqrt{1-\eta}(1-\hat{n}_j),
\nonumber \\
\hat{M}^{1}_j&=&\sqrt{\eta}(1-\hat{n}_j). \nonumber 
\end{eqnarray}
The local probabilities of two possible readouts are given by $p^1_j=\eta \langle \psi(t)|(1-\hat{n}_j)|\psi(t)\rangle$, 
and $p^0_j=1-p^1_j$. Without loss of generality, 
we choose the initial state as Neel state i.e. $|\psi(t=0)\rangle=\Pi_{j=1}^{L/2}\hat{c}_{2j-1}^{\dag}|0\rangle$. 
However, we have checked the robustness of our results by investigating different 
initial states (see Appendix~\ref{appendixI}).
We choose $\Delta t=10^{-2}$.  We  define a parameter $g=\eta/\Delta t$,
In the limit $g<<1$ (i.e. $\eta<<\Delta t$), the occurrence of $\hat{M}^{1}_j$ takes place with extremely low possibility and also, $\hat{M}^{0}_j$ is close to an identity operator. That implies that in this limit the dynamics is close to unitary with extremely rare measurements. On the other hand, 
for $g>>1$, the measurements occur very frequently. In this regime, one may expect quantum Zeno effect, for which the final state should be close to the initial state.
The question we wish to address here is that if there is a critical value of measurement strength $g_c$ above which the many-body quantum Zeno transition occurs? And if so, then what is the nature of this transition? Our main results are reported in Fig.~\ref{fig0}, where we have identified the phase diagram of many-body quantum Zeno transition as a function of interaction strength $V$ and measurement strength $g$.
We use entanglement entropy and purity as two diagnostics for characterizing such transition. In the subsequent sections we use discuss the details.
 
\begin{figure}
    \centering
    \includegraphics[width=0.48\textwidth]{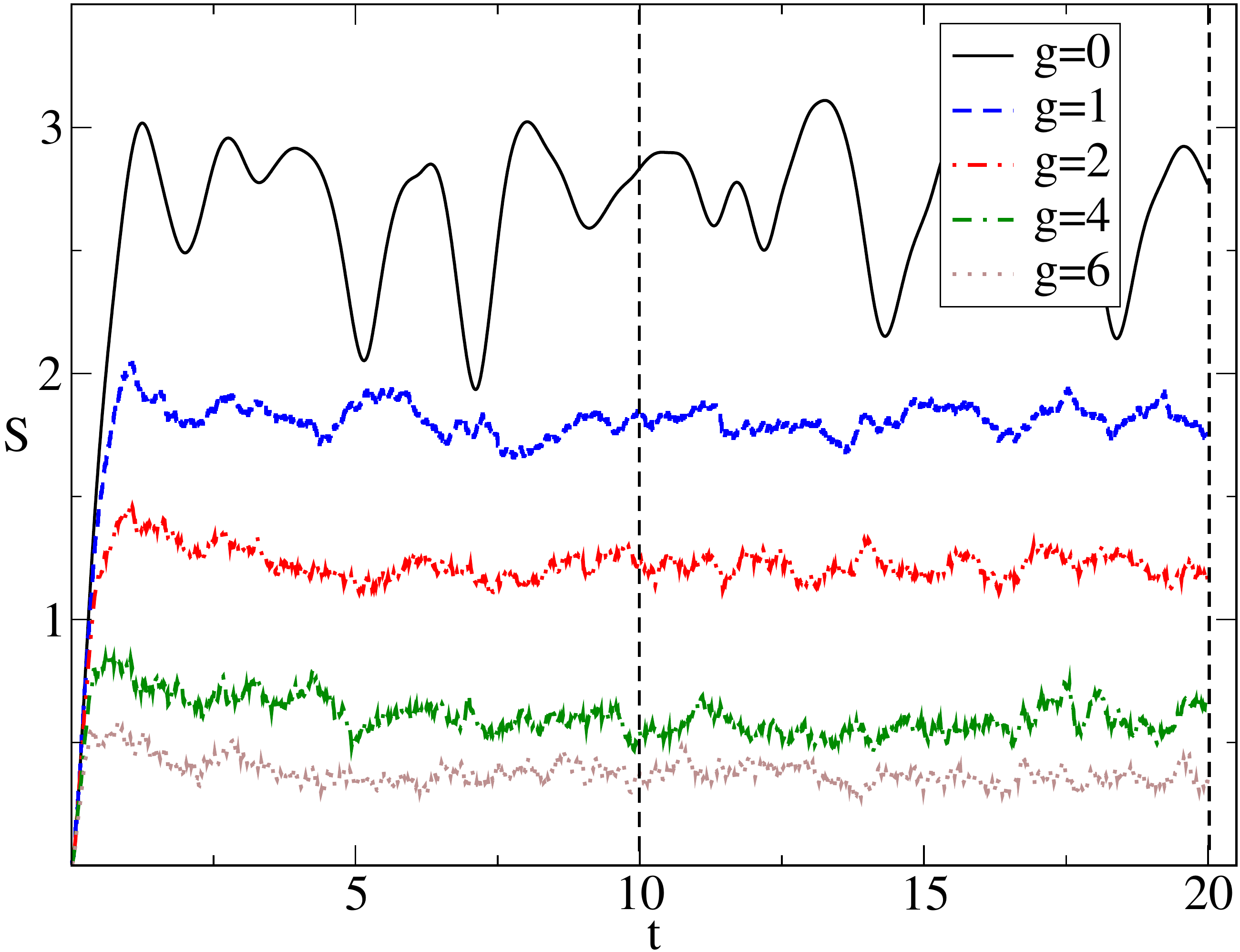}
    \caption{Variation of $S(t)$ with $t$ for different values of $g$ for $L=8$ and $V=1$. Vertical dashed lines represent the time window with $T_1=10$ and $T_2=20$, which we have chosen to calculate the long time average of $S(t)$.}
    \label{fig1}
\end{figure}

\begin{figure}
    \centering
    \includegraphics[width=0.48\textwidth]{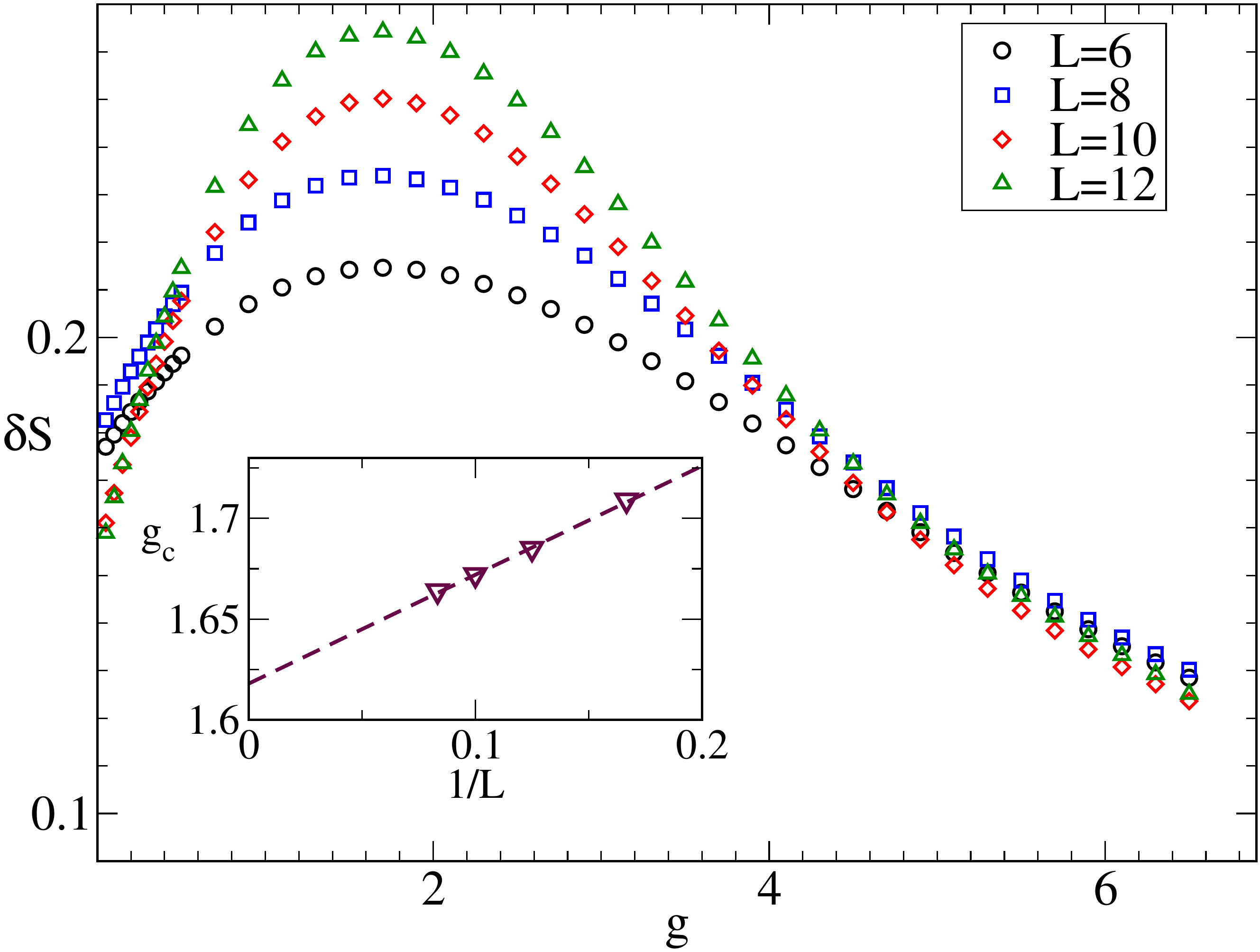}
    \caption{Variation of the fluctuation of the long time average of $S(t)$ with $g$ for different values of $L$ for $V=1$. Inset shows the variation of $g_c$ with $1/L$. Here $g_c$ is identified  as the value of $g$, for which $\delta S$ is maximum. Dashed lines corresponds to the best fit, i.e. $g_c(L)=1.62+0.54/L$.    }
    \label{fig3}
\end{figure}

\begin{figure}
    \centering
    \includegraphics[width=0.48\textwidth]{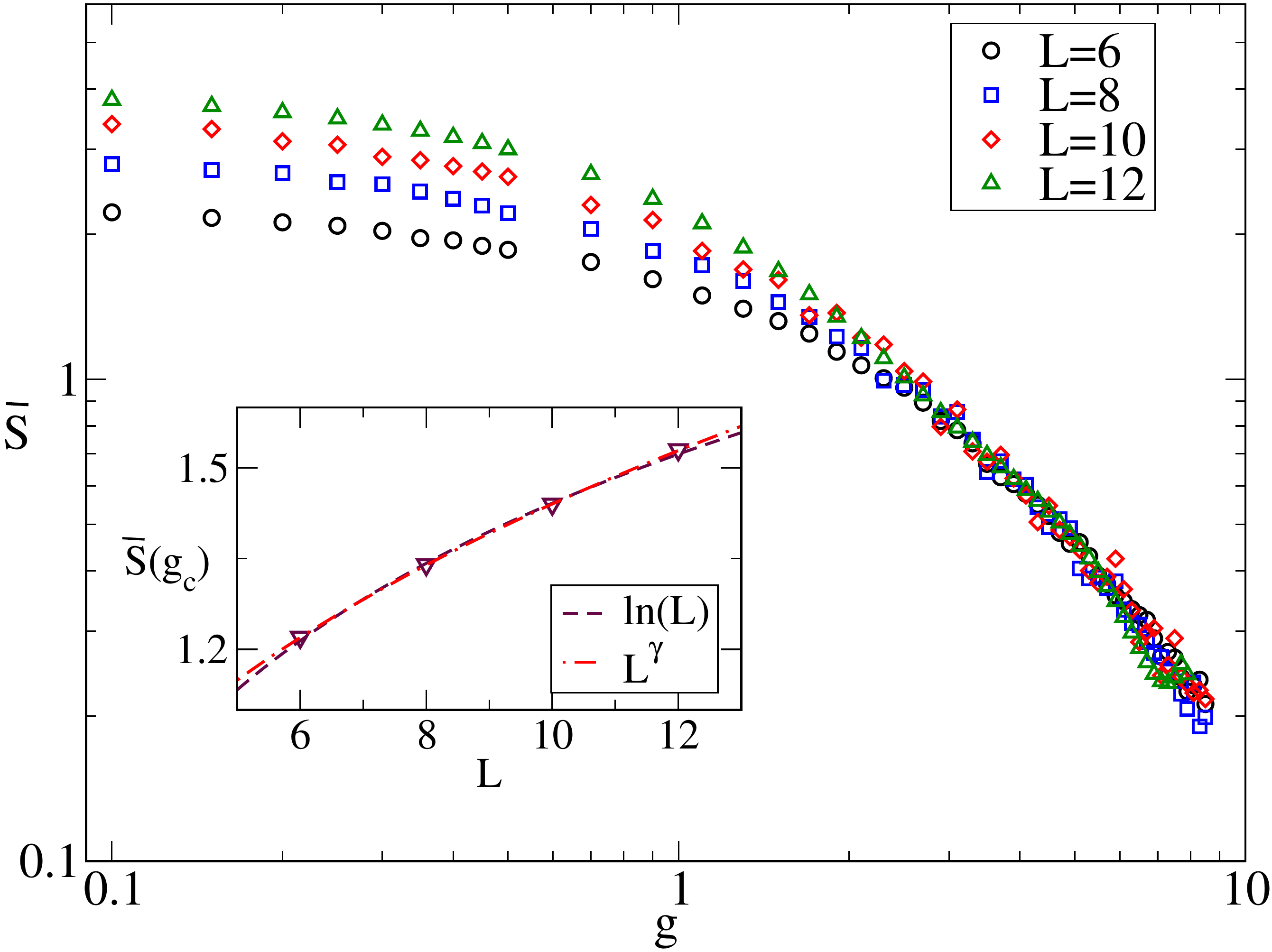}
    \caption{Variation of the long time average of $S(t)$ with $g$ for different values of $L$ for $V=1$. Inset shows the variation of $\overline{S}(t)$ obtained at $g_c$ with $L$. Dashed line  and dashed-dot line correspond to $\overline{S}(g_c)\approx 0.445\ln L +0.417$
    and $\overline{S}(g_c)\approx L^{0.326}$, respectively. }
    \label{fig2}
\end{figure}
\begin{figure}
    \centering
    \includegraphics[width=0.48\textwidth]{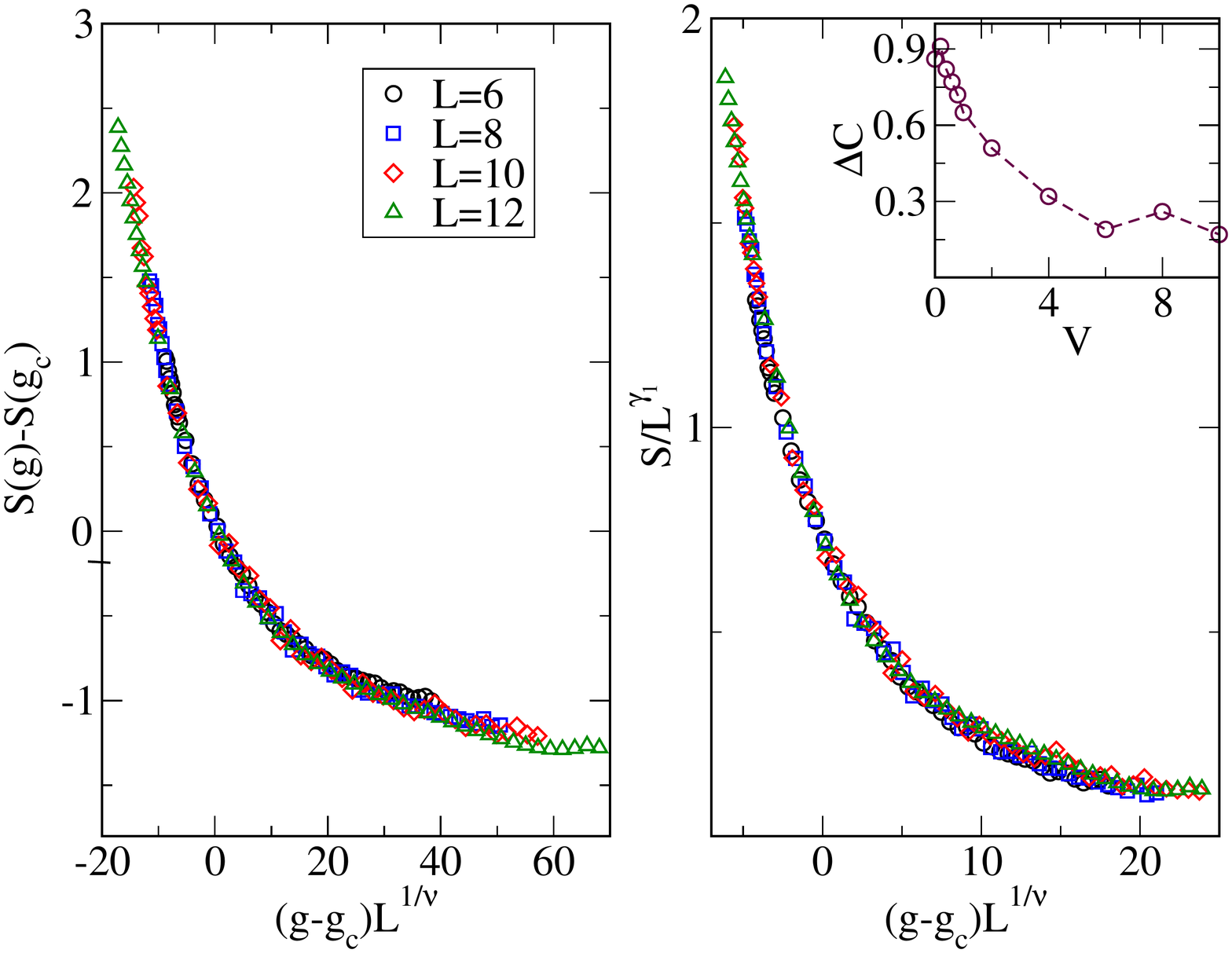}
    \caption{Finite-size entanglement scaling and data collapse. Left panel $g_c=1.63$,  and $\nu=1.05$. and for right panel $g_c=1.65$, $\gamma_1=0.3$, and $\nu=1.84$. Inset of right panel figures shows the variation of $\Delta C$ with $V$.}
    \label{fig2_2}
\end{figure}

\section{Entanglement dynamics}
We study the entanglement dynamics for our protocols. In the limit $g=0$, the dynamic is purely unitary. However, for $g>0$ given that in each small time intervals $\Delta t$, unitary evolution is followed by a measurement i.e. $|\psi(t+\Delta t)\rangle=\hat{M}\hat{U}|\psi(t)\rangle$, the dynamics does not remain unitary. Hence, the state is required to be normalized in each step. 
The reduced density matrix $\rho_A$ of a finite  subsystem A of length  $L/2$ is defined as $\rho_A = \Tr_B|\psi(t)\rangle \langle \psi(t)|$, where the trace is over the degrees of freedom of the complement B of A.  Here we only restrict ourselves to one of the most useful entanglement measures, the von Neumann (entanglement) entropy $S=-\Tr[\rho_A\ln \rho_A]$. It is well known that in the limit  $g=0$,  when the initial Neel state is evolved under the unitary evolution governed by the Hamiltonian $\hat{H}$, $S(t)$
grows linearly with time $t$ for short time, while in the long time limit it saturates 
to  the thermodynamic entropy densities of the generalized Gibbs ensemble (GGE),  which is extensive in subsystem size~\cite{alba1,alba2}. 

Figure.~\ref{fig1} shows the variation of $S(t)$ with $t$ for different values of $g$ for a fixed $L=8$. We repeat our stochastic measurement protocols for $N=1000$ times and the data presented  in Fig.~\ref{fig1} is averaged over different realizations. 
We find that given that the initial state is a zero entangled product state, $S(t)$ grows for a short time, and then it saturates.  The long time saturation value of $S(t)$ decreases as we increase $g$. This result can be anticipated as one may expect many-body quantum Zeno effect in the limit $g>>1$. In this limit the final state should be close to the initial Neel state, which is a product state.
We evaluate the long time average value of $S(t)$ i.e.
$\overline{S(t)}=\frac{1}{T_2-T_1}\int_{T_1}^{T_2}{S(t)}$. We choose $T_2=20$, and $T_1=10$. 

Then we  investigate the fluctuation of $\overline{S(t)}$. Given that we have repeated our protocols $N=1000$ times, we calculate the standard deviation 
of $\overline{S(t)}$ i.e. $\delta S= \sqrt{\langle\overline{S}^2\rangle- (\langle\overline{S}\rangle)^2}$, where $\langle  \rangle$ stands for averaging  over $N$ times \cite{luitz}. The standard deviation of the entanglement entropy displays a maximum at the many-body localization phase transition. We find  analogous features even for many-body Zeno transition in Fig.~\ref{fig3}, and we identify  the value of $g$ for which $\delta S$ is maximum as a transition point for a given system size. In the inset of Fig.~\ref{fig3}, we plot the critical value of $g$, i.e. $g_c$ as a function of $1/L$, and the extrapolated data gives a rough estimation of $g_c$ in the thermodynamics limit, which we have found to be $1.62$.

Next, in Fig.~\ref{fig2}, we plot the variation of $\overline{S(t)}$ with $g$ for different values of $L$. Remarkably, we find that above some value of $g$, $\overline{S(t)}$ obeys area law i.e. remains independent of $L$, while for small values of $g$,  $\overline{S(t)}$ increases with 
the system size. This results indicates a clear signature of many-body Zeno transition beyond a critical measurement strength $g_c$. 
However, in order to understand a phase-transition
from a finite size results, one needs to perform finite size scaling analysis. On the other hand in order to determine the correct scaling functional form, it is important to know, how $\overline{S}$ scales with $L$ at $g_c$. In the inset of Fig.~\ref{fig2}, we plot $\overline{S}(g_c)$ with $L$, where $g_c$ is identified as before i.e. the value $g$ for which $\delta S$ is maximum for a given $L$. The data suggests that $\overline{S}(g_c)$ can be fitted reasonably well using both $\ln L$ and $L^{0.32}$. Due to the system size limitations, it is almost impossible to conclude definitely that  whether $\overline{S}(g_c)$ scales as 
$\ln L$ or $L^{0.32}$. 

In order to obtain data collapse, we use the following scaling ans{\" a}tze: 1) $\overline{S}=L^{\gamma_1}f[(g-g_c)L^{1/\nu}]$ ~\cite{Chen20}, and
2) $\overline{S}-\overline{S}(g_c)=f[(g-g_c)L^{1/\nu}]$ ~\cite{luitz,Fuji20}, where $f[]$ can be some arbitrary function. In the Fig.~\ref{fig2_2} we show the data collapse using both the ans{\" a}tze i.e. in the right panel we  re-scaled $\overline{S}$ by $\overline{S}/L^{\gamma_1}$ and $g$ by $(g-g_c)L^{1/\nu}$  and in the left panel we plot $\overline{S}-\overline{S}(g_c)$ in the y axes and re-scaled in $x$ axes $g$ by $(g-g_c)L^{1/\nu}$. 
In order to find the best fit values of the scaling exponents and $g_c$ from the data collapse, we use the cost function approach which was introduced in Ref~\cite{lev1}
recently. The cost function is defined for a quantity $X$ that consists of $N_p$  values at different $g$ and $L$ as, 
\begin{eqnarray}
C_X=\frac{\sum_{j=1}^{N_p-1}|X_{j+1}-X_j|}{max(X_j)-min(X_j)} -1.\nonumber 
\end{eqnarray}
Then we sort all $N_p$ values of $X_j$ according to
non-decreasing values of $L$ sgn$[g-g_c]|g-g_c|^{\nu}$. In the case of ideal data collapse 
$\sum_{j=1}^{N_p-1}|X_{j+1}-X_j|=max(X_j)-min(X_j)$, hence $C_X=0$. However, otherwise 
$C_X>0$. One is reminded here of the concept of visibility in interference phenomena \cite{Jaege95}. For the 1st ansatz,  we choose $X=\overline{S}/L^{\gamma_1}$, and find the values of $\gamma_1$, $\nu$ and $g_c$, for which the cost function gets minimized. 
The best data collapse is obtained for $\gamma_1=0.3$, the scaling exponent $\nu=1.84$, and critical measurement strength  $g_c=1.65$. Since the best fit value obtained for $\gamma_1$ is $<1$, that makes quantum Zeno transition different compared to many-body localization (MBL) transition. While in the MBL transition near the critical point, the entanglement entropy shows volume law scaling i.e. $\gamma=1$ ~\cite{luitz,lev1}, here in the quantum Zeno transition the entanglement entropy scaling obeys sub-volume law. We do similar exercises for 2nd ansatz as well, but in two steps. In the first step the coarse-grained value of $g_c$, obtained from logarithmic extrapolation, as discussed in the inset of Fig.~4, is used in  $\overline{S}(g_c)$. The fine grained value of the critical point and other scaling exponents are extracted via cost funtion minimization in the final step. The obtained values are $g_c=1.63$ and $\nu=1.05$.  

However, the question which is still remain unanswered is that which of these two ansatzs gives better data collapse. For that we plot the difference between cost function values obtained after carried out minimization process using both the ansats. In the inset of Fig.~\ref{fig2_2} (right panel), we show the variation of $\Delta C=C_1-C_2$ with the interaction strength $V$ (where $C_1$ ($C_2$) is  the value of cost function obtained for 1st (2nd) ansatz). We find that $\Delta C > 0$ and that indicates
that 2nd ansatz is much better suited for our data.

%Furthermore we also investigate the fluctuation of %$\overline{S(t)}$. Given that 
%we have repeated our protocols $N=1000$ times, we calculate the standard deviation 
%of $\overline{S(t)}$ i.e. $\delta S= \sqrt{\langle\overline{S}^2\rangle- (\langle\overline{S}\rangle)^2}$, where $\langle  \rangle$ stands for averaging  over $N$ times. As  argued in Ref.~\cite{luitz}, the
%standard deviation of the entanglement entropy displays a
%maximum at the many-body localization phase transition. We find  analogous features even for many-body Zeno transition 
%in Fig.~\ref{fig3}. 
%A scaling collapse of the form
%(see the inset of Fig.~\ref{fig3}). $\delta S=L^{\gamma_1}f[(g-g_c)L^{1/\nu}]$ is also observed, where the scaling exponent $\nu$ and critical measurement strength $g_c$ 
%found using the minimizing cost function  to be very close the one obtained earlier. 

\begin{figure}
    \centering
    \includegraphics[width=0.48\textwidth]{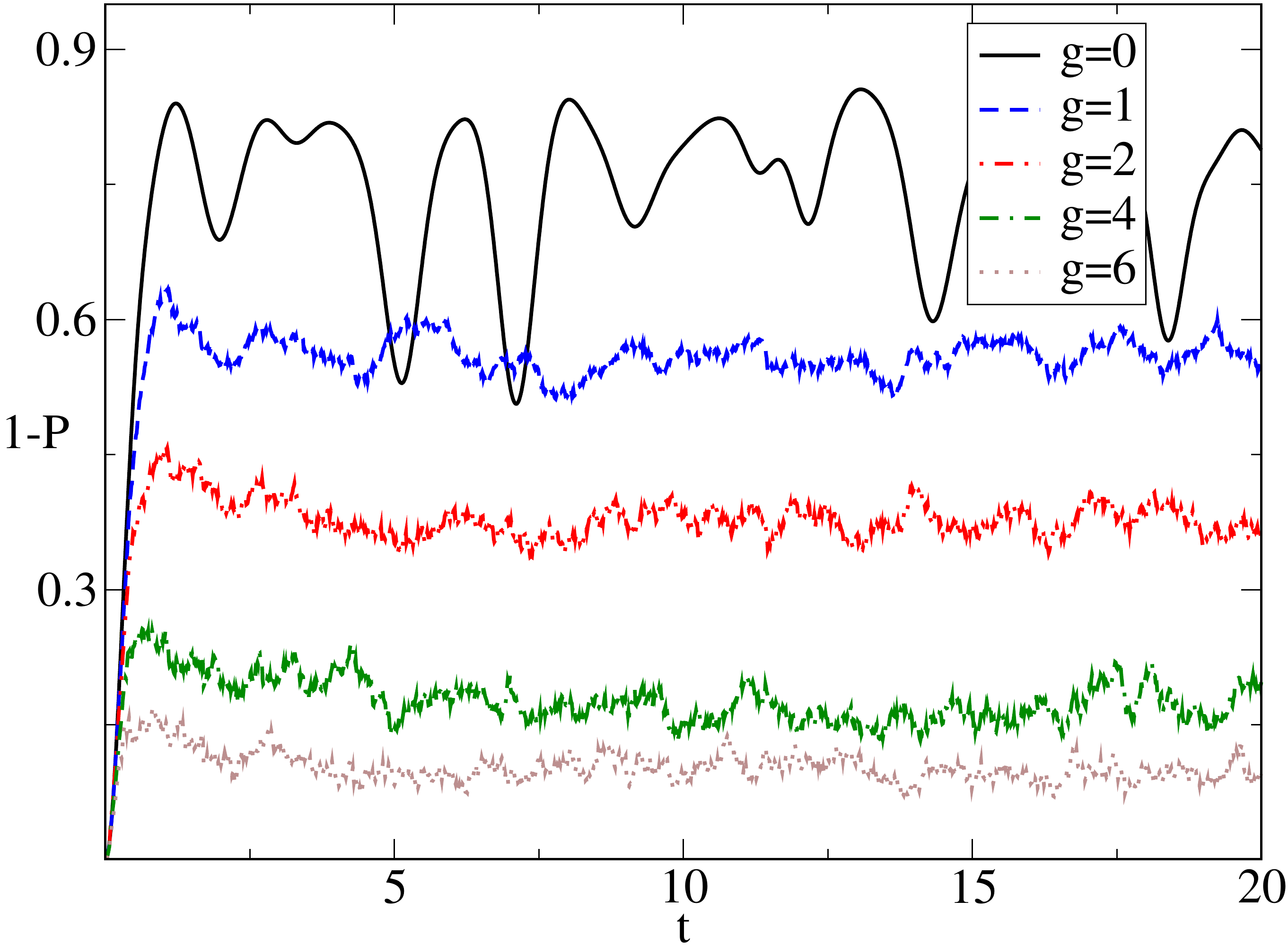}
    \caption{Variation of the mixedness $1-P(t)$ with $t$ for different values of $g$ for $L=8$ and $V=1$. Vertical dashed lines represent the time window with $T_1=10$ and $T_2=20$, which we have chosen to calculate the time average of $1-P(t)$.}
    \label{fig4}
\end{figure}

\begin{figure}
    \centering
    \includegraphics[width=0.48\textwidth]{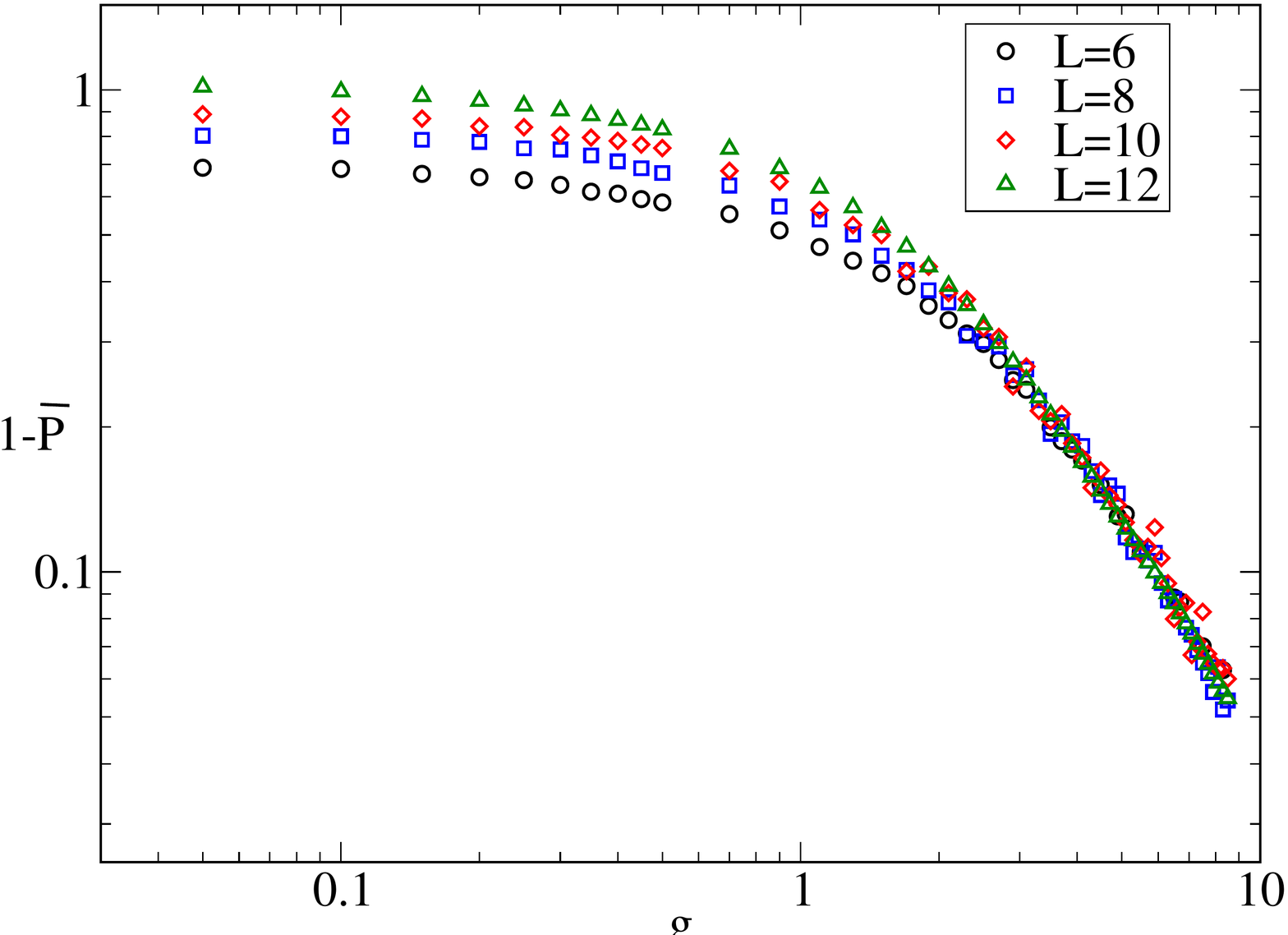}
    \caption{Varation of the long time average of $1-P(t)$ with $g$ for different values of $L$ for $V=1$.  }
    \label{fig5}
\end{figure}

\begin{figure}
    \centering
    \includegraphics[width=0.48\textwidth]{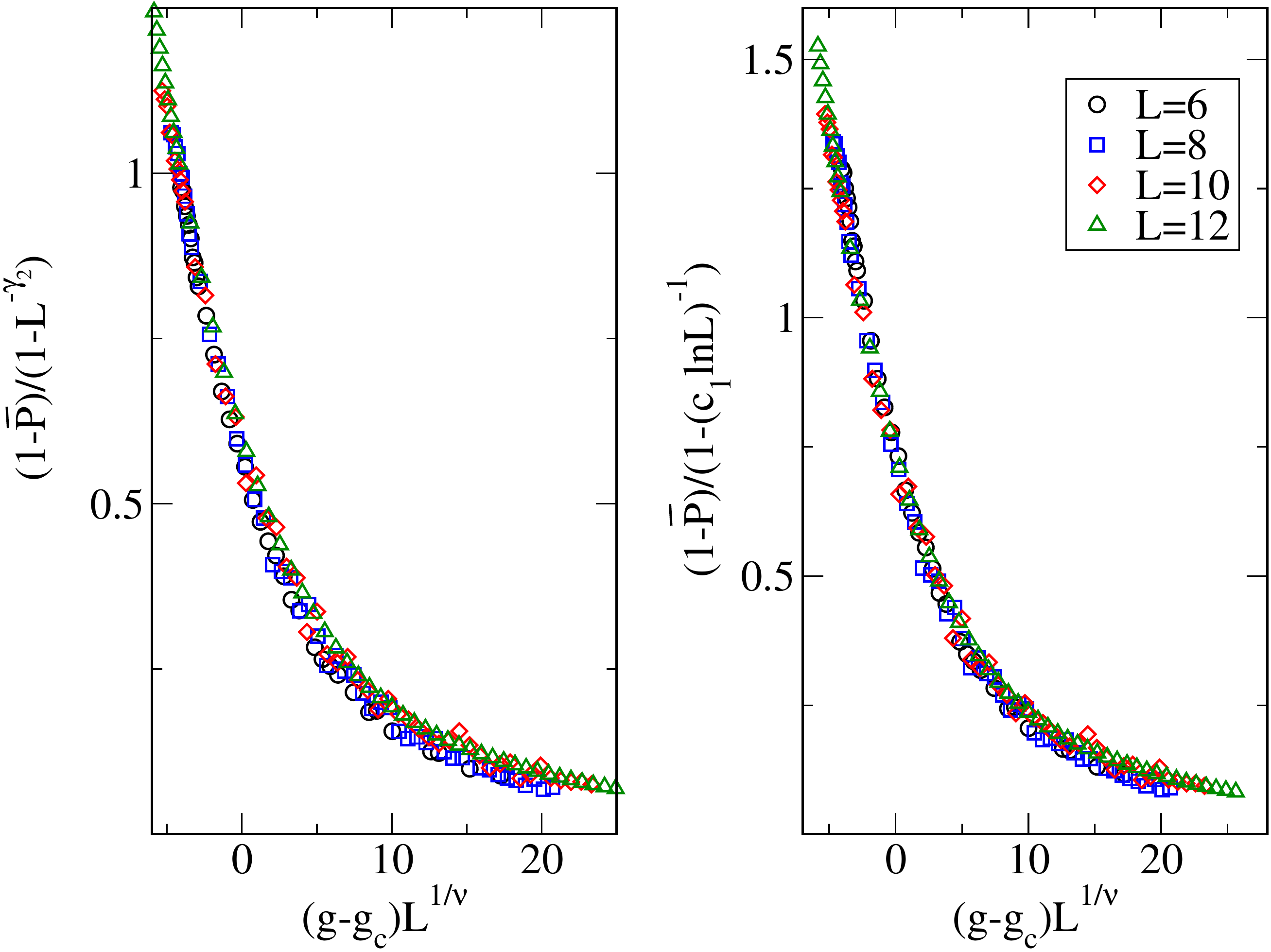}
    \caption{Left panel shows the data collapse using 1st ansatz  where $g_c=1.62$, $\gamma_2=0.68$, and $\nu=1.89$. Right panel shows the data collapse using 2nd ansatz  where $g_c=1.62$, $c_1=0.92$, and $\nu=1.87$. }
    \label{fig5_1}
\end{figure}

\section{Purity}
Next, we use the purity of the reduced density matrix $\rho_A$ as another diagnostic to
characterize the Zeno transition. The purity $P$ is defined as $P=\Tr\rho_A^{2}$.
Given that our initial Neel state remains pure even after bi-partition, one expects that for $g>>1$ (deep into many-body Zeno phase)  $P \simeq 1$. However, in the limit $g<<1$, one expects the time evolved state after bi-partition 
will be mixed, and hence  $\Tr\rho_A^{2} << 1$. 

Figure.~\ref{fig4} shows the variation the mixedness which is defined as $1-P$ of the reduced density matrix as a function of time for different values of measurement strength $g$. As expected, with the increase of $g$, the long-time values of $1-P$ decreases. 
In the same spirit of the entanglement dynamics analysis, we again evaluate long time average of mixedness of a time evolved state i.e. $\overline{1-P(t)}=\frac{1}{T_2-T_1}\int_{T_1}^{T_2}{(1-P(t))}$. We choose $T_2=20$, and $T_1=10$ which is denoted with the vertical dashed line in Fig.~\ref{fig4}. Next, we focus 
on the variation of $1-\overline{P(t)}$ with $g$ for different values of $L$
as shown in Fig.~\ref{fig5}. While we find that for large enough value of $g$, the mixedness does not change with $L$, but for small values of $g$ it increases with increasing $L$. Since the $\overline{1-P(t)}$ is always bounded between $0$ and $1$, 
we assume that near the transition point  $\overline{1-P(t)}$
can be described by either  $\overline{1-P(t)}=(1-L^{-\gamma_2})h[(g-g_c)L^{1/\nu}]$ (ansatz 1) or $\overline{1-P(t)}=(1-(c_1\ln L)^{-1})h[(g-g_c)L^{1/\nu}]$ (ansatz 2), where $h[]$ 
can be some arbitrary function such that $\overline{1-P(t)}$ remains $[0,1]$. 
In the Fig.~\ref{fig5_1} we show the data collapse by re-scaling $1-\overline{P}$ by $(1-\overline{P})/(1-L^{-\gamma_2})$ and $g$ by $(g-g_c)L^{1/\nu}$ (in the left panel) and by re-scaling $1-\overline{P}$ by $(1-\overline{P})/(1-(c_1\ln L)^{-1})$ and $g$ by $(g-g_c)L^{1/\nu}$ (in the right panel). The best fit values of different scaling exponents and $g_c$ for the data collapse are obtained once again minimizing  the cost function, and $\gamma_2=0.68$,   $\nu=1.89$ ($1.87$), and the critical measurement strength  $g_c=1.62$ ($1.62$) are obtained as the best fit values using 1st (2nd) ansatz . 
\begin{table}
\begin{center}
\begin{tabular}{ |c|c||c|c|c|c|c| } 
 \hline
 Method & Ansatz & $g_c$ & $\nu$ & $\gamma_1$ &  $\gamma_2$ &$C_X$\\ 
 \hline
 $\overline{S}$ & 1st &1.65 & 1.84 & 0.30 & - &2.156\\ 
 \hline
  $\overline{S}$ & 2nd &1.63 & 1.05 & -& - &1.266\\ 
\hline
 $\delta S$ & & 1.62 & - & - & -& \\ 
 \hline
 $\overline{P}$& 1st & 1.62 & 1.89 &- & 0.68 & 2.345 \\
 \hline
  $\overline{P}$& 2nd & 1.62 & 1.87 &- & - & 1.55 \\
\hline
\end{tabular}
\caption{Comparison of the values of different scaling exponents and critical measurement strength obtained from different methods. }
 \label{table1}
\end{center}
\end{table}
\section{Conclusions}
In this paper, we explored many-body quantum Zeno effect in an one-dimensional interacting fermionic system subjected to repeated projective measurements. By tuning the measurement strength, we demonstrated measurement-induced dynamical phase transition due to the competition between unitary time evolution and projective measurements. The transition occurs between two distinct phases -  (sub)-volume-law obeying entanglement entropy in the long-time evolved steady states in the limits of weak interactions and area-law obeying long-time evolved steady states in the limits of strong interactions (quantum Zeno phase). We gain further insights by investigating the fluctuations of entanglement entropy. Following the analysis on entanglement entropy, we turned our focus to the purity of the reduced density matrices. While, in the Zeno phase the purity remain independent of the system size, the purity grows with system size beyond the transition point. 

Next we performed careful scaling analyses for systems with finite interactions. The scaling analysis is carried out by employing the measures under study, i.e. entanglement entropy, purity and their fluctuations. Information regarding the transition points and scaling exponents are extracted via an unbiased numerical strategy of cost function minimization that ensures the quality of finite-size data collapse. We discuss finite-size scaling ans{\" a}tze proposed previously context of many-body Zeno transition. 
Cost function minimization approach provides a powerful means for examining their quality. Finally, we provide the phase diagram involving quantum Zeno transition for the interacting system.

In future it will be interesting devise a concrete formalism for an eﬀective description via a non-Hermitian Hamiltonian, which has only been non-rigorously suggested in context of other systems \cite{Halliwell10,Echanobe08,Halliwell09,Halliwell10}. Another general interest is to build a  quasi-particle description for the dynamics of entanglement entropy~\cite{vc2017} for systems under repetitive projective measurements.

\section{Acknowledgements}
RM acknowledges the support of DST-Inspire fellowship, by the Department of Science and Technology, Government of India.  US acknowledges partial support from the Department of Science and Technology, Government of India through the QuEST grant (Grant No.~DST/ICPS/QUST/Theme-3/2019/120). Super computing Facilities at IIT (BHU) were used to perform numerical computations.

\section{Appendix I}\label{appendixI}
\begin{figure}[h!]
    \centering
    \includegraphics[width=0.48\textwidth]{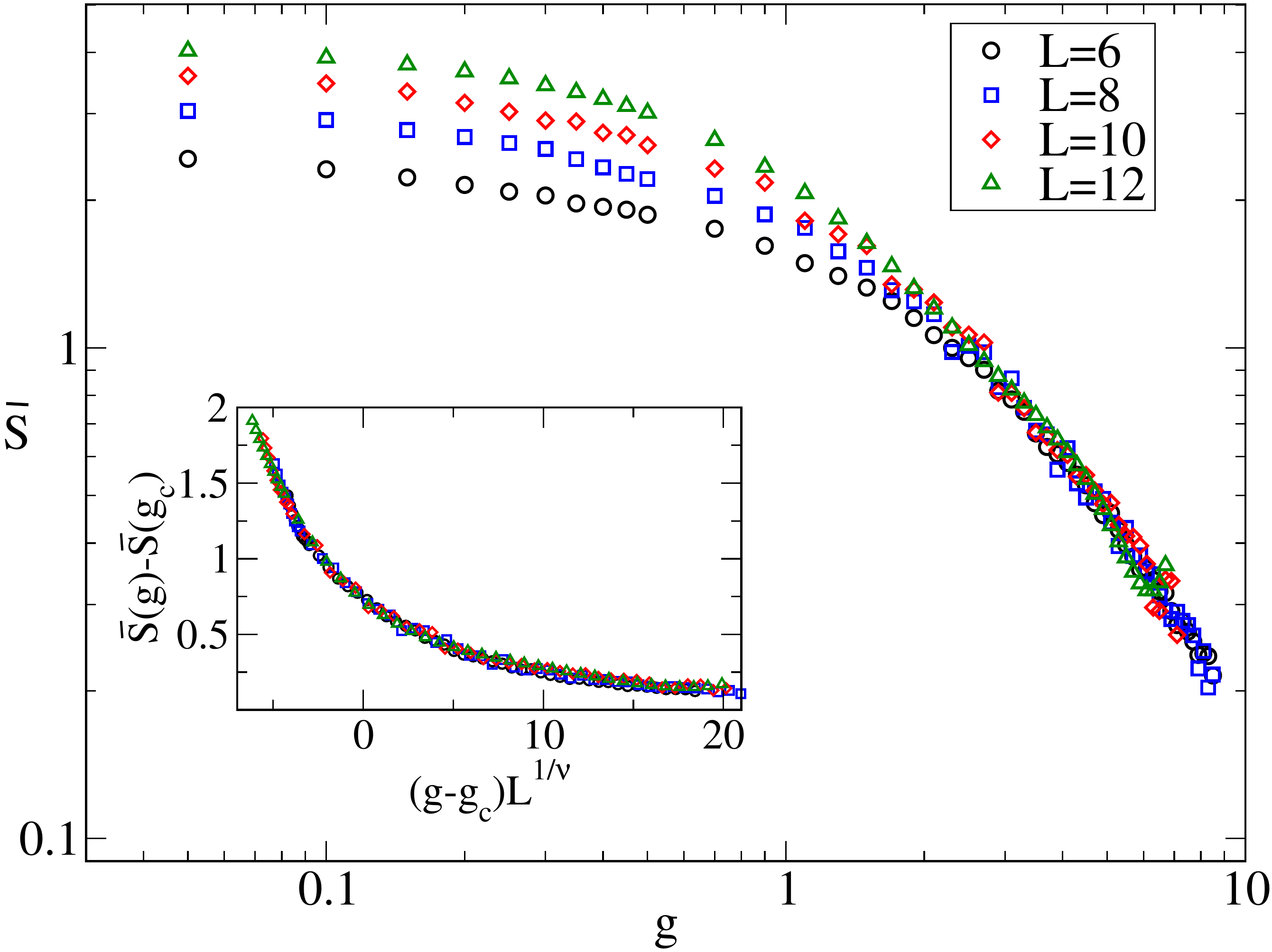}
    \caption{Variation of the long time average of $S(t)$ with $g$ for different values of $L$ for $V=1$. Inset shows the data collapse, where $g_c=1.62$, and $\nu=1.11$. }
    \label{fig6}
\end{figure}

In this appendix we study the many-body Zeno transition for the initial state $|\psi(t=0)\rangle=\Pi _{i=1}^{L/2}\hat{c}_i^{\dag}|0\rangle$. As mentioned in the main text (where we have chosen the Neel state as an initial state), we first calculate the entanglement dynamics  for different values of measurement strength $g$. Then we evaluate the long-time average $\overline{S(t)}$. Next, we investigate $\overline{S(t)}$ with $g$ 
for different system sizes. Figure.~\ref{fig6} shows the variation of $\overline{S}$ with 
$g$ for $L=6$, 8, 10, 12. Inset of Fig.~\ref{fig6} once again shows the data collapse using ansatz 2, 
surprisingly $g_c$, and $\nu$ obtained from the best fit is very close to the 
one reported in the main text for the Neel state. It proves the robustness of our finding. 

\begin{thebibliography}{100}

\bibitem{Chou05} C. W. Chou et al., \emph{Measurement induced entanglement for excitation stored in remote atomic ensembles}, Nature {\bf 438}, 828 (2005).
\bibitem{Roch14} N. Roch et al., \emph{Observation of Measurement-Induced Entanglement and Quantum Trajectories of Remote Superconducting Qubit}, Phys. Rev. Lett. {\bf 112}, 170501 (2014). 
\bibitem{Poop05} M. Popp, F. Verstraete, M. A. Martin-Delgado and J. I. Cirac, \emph{Localizable entanglement}, Phys. Rev. A {\bf 71}, 042306 (2005).
\bibitem{Sadhukhan17} D Sadhukhan, SS Roy, AK Pal, D Rakshit, A Sen and U Sen, \emph{Multipartite entanglement accumulation in quantum states: Localizable generalized geometric measure}, Physical Review A 95, 022301 (2017).


\bibitem{Raussendorf} R. Raussendorf and H. J. Briegel, Phys. Rev. Lett. {\bf 86}, 5188 (2001).
\bibitem{Sorensen03} A. S. S{o}rensen and K. M{o}lmer, \emph{Measurement Induced Entanglement and Quantum Computation with Atoms in Optical Cavities}, Phys. Rev. Lett. 91, 097905 (2003).
\bibitem{Brigel09} H. J. Briegel, D. Browne, W. Dur, R. Raussendorf, and M. van den Nest, \emph{Measurement-based quantum computation}, Nat. Phys. {\bf 5}, 19 (2009).

\bibitem{Koshino05} K. Koshino and A. Shimizu, \emph{Quantum Zeno effect by general measurements}, Phys.Rept. {\bf 412}, 191 (2005).
\bibitem{Wiseman09} H. M. Wiseman and G. J. Milburn, \emph{Quantum measurement and control}, Cambridge university press (2009). 


\bibitem{Misra77} B. Misra and E. C. G. Sudarshan, \emph{The Zeno's paradox in quantum theory}, J. Math. Phys. {\bf 18}, 756 (1977).

\bibitem{Suntajs20} J. Suntajs, J. Bonca,2, 1 T. Prosen, and L. Vidmar, \emph{Ergodicity Breaking Transition in Finite Disordered Spin Chains}, Phys. Rev. B {\bf 102}, 064207 (2020).

\bibitem{KOfman00} A. G. Kofman and G. Kurizki, \emph{Acceleration of quantum decay processes by frequent observations}, Nature {\bf 405}, 546 (2000).
\bibitem{Facchi02} P. Facchi and S. Pascazio, \emph{Quantum Zeno subspaces}, Phys. Rev. Lett. {\bf 89}, 080401 (2002).
\bibitem{Maniscalco08} S. Maniscalco, F. Francica, R. L. Zaffino, N. Lo Gullo and F. Plastina, \emph{Protecting entanglement via the quantum Zeno effect}, Phys. Rev. Lett. {\bf 100}, 090503 (2008).
\bibitem{Facchi10} P. Facchi and M. Ligab{\'o}, \emph{Quantum Zeno effect and dynamics}, J. Phys. A: Math. Theor. {\bf 51}, 022103 (2010).
\bibitem{Militello11} B. Militello, M. Scala and A. Messina, \emph{Quantum Zeno subspaces induced by temperature}, Phys. Rev. A {\bf 84}, 022106 (2011).
\bibitem{Raimond12} J. M. Raimond \emph{et al.}, \emph{Quantum zeno dynamics of a field in a cavity}, Phys. Rev. A {\bf 86}, 032120 (2012).
\bibitem{Wang13} S. C. Wang, Y. Li, X. B. Wang and L. C. Kwek, \emph{Operator quantum Zeno effect: Protecting quantum information with noisy two-qubit interactions}, Phys. Rev. Lett, {\bf 110}, 100505 (2013).


 
\bibitem{Stannigel} K. Stannigel  \emph{et al.}, \emph{Constrained dynamics via the Zeno effect in quantum simulation: implementing non-abelian lattice gauge theories with cold atoms}, Phys. Rev. Lett. {\bf 112}, 120406 (2014).
\bibitem{Froml19} H. Fröml et al., \emph{Fluctuation Induced Quantum Zeno Eﬀect}, Phys. Rev. Lett. {\bf 122}, 040402 (2019).
\bibitem{Froml20} H. Fröml et al. \emph{Ultracold quantum wires with localized losses: Many-body quantum Zeno eﬀect}, Phys. Rev. B 101, 144301 (2020).
\bibitem{Syassen08} N. Syassen et al., \emph{Strong Dissipation Inhibits Losses and Induces Correlations in Cold Molecular Gases}, Science {\bf 320}, 1329 (2008).
\bibitem{Biella19} A. Biella and M. Schir{\'o}, \emph{Many-body quantum Zeno eﬀect and measurement-induced subradiance transition}, Phys. Rev. Lett. {\bf 122}, 040402 (2019).



\bibitem{Amico08} L. Amico, R. Fazio, A. Osterloh, and V. Vedral, Entanglement in many-body systems, Rev. Mod. Phys. {\bf 80}, 517 (2008).
\bibitem{Eisert10} J. Eisert, M. Cramer, and M. B. Plenio, Colloquium: Area laws for the entanglement entropy, Rev. Mod. Phys. {\bf 82}, 277 (2010).

\bibitem{Islam15} R. Islam, R. Ma, P. M. Preiss, M. E. Tai, A. Lukin, M. Rispoli, and M. Greiner, \emph{Measuring entanglement entropy in a quantum many-body system}, Nature, {\bf 528}, 7580 (2015).
\bibitem{Kaufman16} A. M. Kaufman, M. E. Tai, A. Lukin, M. Rispoli, R. Schittko, P. M. Preiss, and M. Greiner, \emph{Quantum thermalization through entanglement in an isolated manybody system}, Science {\bf 353}, 794 (2016).

\bibitem{Chen18} Y. Li, X. Chen, and M. P. A. Fisher, \emph{Quantum zeno eﬀect and the many-body entanglement transition}, Phys. Rev. B {\bf 98}, 205136 (2018). 
\bibitem{Skinner19} B. Skinner, J. Ruhman, and A. Nahum, \emph{Measurement-induced phase transitions in the dynamics of entanglement}, Phys. Rev. X {\bf 9}, 031009 (2019). 
\bibitem{Li19} Y. Li, X. Chen, and M. P. A. Fisher, \emph{Measurement-driven entanglement transition in hybrid quantum circuits}, Phys. Rev. B {\bf 100}, 134306 (2019). 
\bibitem{Gullans19} M. J. Gullans and D. A. Huse, \emph{Dynamical puriﬁcation phase transitions induced by quantum measurements}, (2019), arXiv:1905.05195 [quant-ph]. 
\bibitem{Gullans20} Michael J. Gullans and David A. Huse, \emph{Scalable probes of measurement-induced criticality}, Phys. Rev. Lett. {\bf 125}, 070606 (2020). 
\bibitem{Jian20} C. -M. Jian, Y. -Z. You, R. Vasseur, and A. W. W. Ludwig, \emph{Measurement-induced criticality in random quantum circuits}, Phys. Rev. B {\bf 101}, 104302 (2020). 
\bibitem{Bao20} Y. Bao, S. Choi, and E. Altman, \emph{Theory of the phase transition in random unitary circuits with measurements}, Phys. Rev. B {\bf 101}, 104301 (2020).
\bibitem{Choi20} S. Choi, Y. Bao, X. -Liang Qi, and E. Altman, \emph{Quantum error correction in scrambling dynamics and measurement-induced phase transition}, Phys. Rev. Lett. {\bf 125}, 030505 (2020). 
\bibitem{Fan20} R. Fan, S. Vijay, A. Vishwanath, and Y. -Z. You, \emph{Self-organized error correction in random unitary circuits with measurement}, arXiv:2002.12385 [cond-mat.stat-mech]. 

\bibitem{Dhar16} S. Dhar and S. Dasgupta, \emph{Measurement-induced phase transition in a quantum spin system}, Phys. Rev. A {\bf 93}, 050103(R) (2016).
\bibitem{Turkeshi20a} X. Turkeshi, R. Fazio, and M. Dalmonte,\emph{Measurement-induced criticality in (2+1)-dimensionalhybrid quantum circuits}, Phys. Rev. B {\bf 102} (2020).
\bibitem{Turkeshi20b} N. Lang and H. P. B{\"u}chler, \emph{Entanglement transition in the projective transverse field ising model}, Phys. Rev. B {\bf 102}, 094204 (2020).
\bibitem{Maimbourg21} T. Maimbourg, D. M. Basko, M. Holzmann and A. Rosso, \emph{Bath-induced Zeno localization in driven many-body quantum systems}, Phys. Rev. Lett. {\bf 126}, 120603 (2021).

\bibitem{Minato21} T. Minato, K. Sugimoto, T. Kuwahara, and K. Saito, \emph{Fate of measurement-induced phase transition in long-range interactions}, arXiv:2104.09118.

\bibitem{Zabalo20} A. Zabalo, M. J. Gullans, J. H. Wilson, S. Gopalakrishnan, D. A. Huse, and J. H. Pixley, \emph{Critical properties of the measurement-induced transition in random quantum circuits}, Phys. Rev. B 101, 060301(R) (2020).

\bibitem{Zabalo21} A. Zabalo, M. J. Gullans, J. H. Wilson, R. Vasseur, A. W. W. Ludwig,S. Gopalakrishnan, D. A. Huse, and J. H. Pixley, \emph{Operator scaling dimensions and multifractality at measurement-induced transitions}, arXiv:2107.03393

\bibitem{Fuji20} Y. Fuji and Y. Ashida, \emph{Measurement-induced quantum criticality under continuous monitoring}, Phys. Rev. B {\bf 102}, 054302 (2020).
\bibitem{Alberton21} O. Alberton, M. Buchhold and S. Diehl, \emph{Entanglement transition in a monitored free fermion chain -- from extended criticality to area law}, Phys. Rev. Lett. {\bf 126}, 170602 (2021).
\bibitem{Chen20} X. Chen, Y. Li, M. P. A. Fisher and A. Lucas, \emph{Emergent conformal symmetry in non-unitary random dynamics of free fermions}, Phys. Rev. Research {\bf 2}, 033017 (2020).
\bibitem{Buchhold21} M. Buchhold, Y. Minoguchi, A. Altland, S. Diehl, \emph{Effective Theory for the Measurement-Induced Phase Transition of Dirac Fermions}, arXiv:2102.08381.
\bibitem{Turkeshi21} X. Turkeshi, A. Biella, R. Fazio, M. Dalmonte and M. Schiro, \emph{Measurement-Induced Entanglement Transitions in the Quantum Ising Chain: From Infinite to Zero Clicks}, Phys. Rev. B {\bf 103}, 224210 (2021).


\bibitem{Nielsen11} M. A. Nielsen and I. L. Chuang, \emph{Quantum Computation and Quantum Information}, Cambridge University Press, New York, NY, USA, 2011.
\bibitem{Wilde13} M. M. Wilde, Quantum Information Theory, Cambridge University Press, New York, NY, USA, 2013.
\bibitem{Kleinmann06} M. Kleinmann, H. Kampermann, T. Meyer and D. Bruss, \emph{Physical Purification of Quantum States}, Phys. Rev. A {\bf 73}, 062309 (2006).
\bibitem{Bera16} A. Bera, A. Kumar, D. Rakshit, R. Prabhu, A. Sen(De), and Ujjwal Sen, \emph{Information complementarity in multipartite quantum states and security in cryptography}, Phys. Rev. A {\bf 93}, 032338 (2016).
\bibitem{Fang20}K. Fang and Z.-W. Liu, \emph{No-go theorems for quantum resource purification}, Phys. Rev. Lett. {\bf 125}, 060405 (2020).

\bibitem{Hume07} D. B. Hume, T. Rosenband, and D. J. Wineland, \emph{High Fidelity Adaptive Qubit Detection through Repetitive Quantum Nondemolition Measurements}, Phys. Rev. Lett. {\bf 99}, 120502 (2007).
\bibitem{Jiang09} L. Jiang et al., \emph{Repetitive Readout of a Single Electronic Spin via Quantum Logic with Nuclear Spin Ancillae}, Science {\bf 326}, 267 (2009).
\bibitem{Ofek16} N. Ofek et al., \emph{Extending the Lifetime of a Quantum Bit with Error Correction in Superconducting Circuits}, Nature {\bf 536}, 441 (2016).
\bibitem{Nakajima19} T. Nakajima et al., \emph{Quantum Non-demolition Measurement of an Electron Spin Qubit}, Nat. Nanotechnol. {\bf 14}, 555 (2019).


\bibitem{lev1} J. Suntajs, J. Bonca, T. Prosen, and L. Vidmar, \emph{Ergodicity breaking transition in finite disordered spin chains}; Phys. Rev. B {\bf 102}, 064207 (2020).


%\bibitem{Takahashi72} M. Takahashi and M. Suzuki, \emph{One-Dimensional Anisotropic Heisenberg Model at Finite Temperatures}, Prog. Theor. Phys. {\bf 48},2187 (1972).

\bibitem{alba1} V. Alba and P. Calabrese, \emph{Entanglement and thermodynamics after a quantum quench in integrable systems}; PNAS {\bf 114}, 7947 (2017).
\bibitem{alba2} V. Alba and P. Calabrese, \emph{Entanglement dynamics after quantum quenches in generic integrable systems}; SciPost Phys. {\bf 4}, 017 (2018).

\bibitem{Jaege95} G. Jaeger, A. Shimony and L. Vaidman, \emph{Two interferometric complementaries}, Phys. Rev. A {\bf 51}, 54 (1995).

\bibitem{luitz} D. J. Luitz, N. Laflorencie, and F. Alet, \emph{Many-body localization edge in the random-field Heisenberg chain}; Phys. Rev. B {\bf 91}, 081103(R) (2015).

\bibitem{Echanobe08} J. Echanobe, A. Del Campo, and J. G. Muga, \emph{ Disclosing hidden information in the quantum Zeno effect: Pulsed measurement of the quantum time of arrival}, Phys.Rev. A {\bf 77}, 032112 (2008).
\bibitem{Halliwell09} J. J. Halliwell and J. M. Yearsley, \emph{Arrival times, complex potentials, and decoherent
histories}, Phys. Rev. A {\bf 79}, 062101 (2009).
\bibitem{Halliwell10} J. J. Halliwell and J. M. Yearsley, \emph{On the relationship between complex potentials andstrings of projection operators}, Jour. Phys.A: Math. Theor. {\bf 43}, 445303 (2010).
\bibitem{vc2017} V. Alba, P. Calabrese,
\emph{Entanglement and thermodynamics after a quantum quench in integrable systems}, 
Proceedings of the National Academy of Sciences 114 (30), 7947 (2017)
\end{thebibliography}
\end{document}